\documentclass[
reprint,
aps,amsmath,
amssymb,
prc,
twocolumn,
superscriptaddress,
longbibliography
]{revtex4-2}

\usepackage{natbib}
\usepackage{graphicx}
\usepackage{dcolumn}
\usepackage{bm}
\usepackage{xcolor}
\usepackage{caption}
\usepackage{subcaption}
\usepackage{siunitx}
\usepackage{makecell}

\usepackage{float}

\usepackage{tabularx}
\usepackage{booktabs}

\begin{document}

\title{Ground State Magnetic Dipole Moment of \textsuperscript{40}Sc}

\author{Robert Powel}
\affiliation{Facility for Rare Isotope Beams, Michigan State University, East Lansing, Michigan 48824, USA}
\affiliation{Department of Physics and Astronomy, Michigan State University, East Lansing, Michigan 48824, USA}

\author{B. Alex Brown}
\affiliation{Facility for Rare Isotope Beams, Michigan State University, East Lansing, Michigan 48824, USA}
\affiliation{Department of Physics and Astronomy, Michigan State University, East Lansing, Michigan 48824, USA}

\author{Jason D. Holt}
\affiliation{TRIUMF, Vancouver, British Columbia, Canada V6T 2A3}

\author{Andrew Klose}
\affiliation{Department of Chemistry and Biochemistry, Augustana University, Sioux Falls, South Dakota 57197, USA}

\author{Kristian K\"onig}
\affiliation{Facility for Rare Isotope Beams, Michigan State University, East Lansing, Michigan 48824, USA}

\author{Jeremy Lantis}
\affiliation{Facility for Rare Isotope Beams, Michigan State University, East Lansing, Michigan 48824, USA}
\affiliation{Department of Chemistry, Michigan State University, East Lansing, Michigan 48824, USA}

\author{Kei Minamisono}
\email{minamiso@nscl.msu.edu}
\affiliation{Facility for Rare Isotope Beams, Michigan State University, East Lansing, Michigan 48824, USA}
\affiliation{Department of Physics and Astronomy, Michigan State University, East Lansing, Michigan 48824, USA}

\author{Takayuki Miyagi}
\affiliation{TRIUMF, Vancouver, British Columbia, Canada V6T 2A3}

\author{Skyy Pineda}
\affiliation{Facility for Rare Isotope Beams, Michigan State University, East Lansing, Michigan 48824, USA}
\affiliation{Department of Chemistry, Michigan State University, East Lansing, Michigan 48824, USA}

\date{\today}

\begin{abstract}
The hyperfine coupling constants of the proton dripline odd-odd $^{40}$Sc nucleus were deduced from the hyperfine spectrum of the $3d4s\,^3\text{D}_2$\,$\leftrightarrow$\,$3d4p\,^3\text{F}^\circ_3$ transition in Sc II, measured by the bunched beam collinear laser spectroscopy technique. The ground state magnetic dipole and electric quadrupole moments were determined for the first time as $\mu$\,=\,$+5.57(4)(2)\,\mu_N$ and $Q$\,=\,$+42(38)(28)\,e^2\;{\rm fm^2}$, respectively. The magnetic moment is well reproduced by the additivity rule with magnetic moments of neighboring odd-even nuclei in the vicinity of the doubly-magic $^{40}$Ca nucleus. An ab-initio multishell valence-space Hamiltonian was also employed to calculate the magnetic moment of $^{40}$Sc, which spans across the $sd$ and $fp$ nuclear shells, where we obtained good agreements.
\end{abstract}

\maketitle 

\section{Introduction}
\label{sec:introduction}
Ground state magnetic dipole moments of radioactive nuclei have played a crucial role in understanding their structure. A magnetic moment is a one body operator acting on a single nuclear state and is complementary to magnetic transition moments, and is sensitive to how valence nucleons are arranged inside a nucleus, namely the nuclear wavefunction. Each of the single particle proton and neutron state in the nuclear shell model has a distinct magnetic moment, the so called Schmidt \cite{sch37} or single-particle value. Any departure from the single-particle value indicates requirements of the configuration mixing of valence protons and neutrons, and/or mesonic effects.  

In the present study the ground state magnetic moment of the $^{40}$Sc($I$ = 4$^-$, $T_{1/2}$ = 182.3 ms) nucleus was determined for the first time using the bunched beam collinear laser spectroscopy technique. The $^{40}$Sc nucleus occurs in the vicinity of the stable doubly-magic $^{40}$Ca with one more proton and one less neutron to the $^{40}$Ca nucleus. The $^{40}$Sc nucleus, however, is situated at the proton drip line, since $^{39}$Sc at one less neutron is known to be unbound against the one proton emission \cite{woo88}. Furthermore, due to the nature of the odd-odd nucleus with nucleon configuration across the $sd$ (one neutron hole in $d_{3/2}$ shell) and $fp$ (one proton in the $f_{7/2}$ shell) major shells and possible valence proton-neutron interactions, it is notoriously challenging to reproduce the experimental magnetic moments by any theoretical calculations. 

In the single particle (sp) model, the magnetic moment of an odd-even, half-integer isospin nucleus is given by the projection theorem and determined by the last unpaired nucleon in the odd nuclei group \cite{sch37} as
%
\begin{equation}\label{eq:spmagneticmom}
\mu_{\rm sp} = I\left(g^l \pm \frac{g^s - g^l}{2l + 1}\right),
\end{equation}
%
where superscripts $l$ and $s$ are for the orbital and spin angular momentum, respectively, and nuclear spin $I$ is given by $I = l \pm 1/2$. The free nucleon g-factors are given by $g^l_p$ = 1, $g^s_p$ = 5.5855, $g^l_n$ = 0 and $g^s_n$ = $-$3.826. It is well known that magnetic moments appear in between these single-particle lines \cite{sug73} for $I = l \pm 1/2$ with one exception of $^9$C due to the charge symmetry breaking \cite{mat96, huh98, uts04}. Magnetic moments of magic nuclei appear closer to these single-particle lines, and open-shell nuclei being away from the lines, indicating a departure from the single particle model and increasing contribution from the nucleon configuration mixing. 

On the other hand the magnetic moment of an odd-odd, integer isospin nucleus is determined in the single particle model by the last unpaired proton and neutron, and can be obtained by a sum of corresponding single-particle moments of neighboring odd-even nuclei \cite {fee49, kli52}. 
%
\begin{align}
\vec{I} &= \vec{I}_p + \vec{I}_n,\\ \nonumber
\mu_I &= \left<IM\left|g_p\vec{I}_p + g_n\vec{I}_n\right|IM\right>_{M = I}.
\end{align}
%
Here, $I$ and $\mu_I$ are the spin and magnetic moment of the odd-odd nucleus, respectively, and $I_p$ ($g_p$) and $I_n$ ($g_n$) are the spin (g-factor) of the neighboring nucleus with unpaired proton and neutron, respectively. The magnetic moment can then be obtained as 
%
\begin{align}\label{eq:additivity}
\mu_I = &\frac{I}{2I_p}\left\{1 + \frac{(I_p - I_n)(I_p + I_n +1)}{I(I +1)}\right\}\mu_p\\ \nonumber
&\frac{I}{2I_n}\left\{1 + \frac{(I_n - I_p)(I_p + I_n +1)}{I(I +1)}\right\}\mu_n,
\end{align}
%
where $\mu_x = g_x I_x$ with the subscript $x$ for $p$ or $n$. We call this relation the additivity rule hereafter in this paper. Many magnetic moments of such odd-odd nuclei are known near stable isotopes \cite{STONE200575}, and compared to single-particle values based on the additivity rule \cite{ach14}. For $^{40}$Sc, $\mu_p$ and $\mu_n$ correspond to magnetic moments of $^{41}$Sc and $^{39}$Ca, respectively. It is of primary interest of the present study whether or not the magnetic moment of proton drip-line $^{40}$Sc can be expressed using know magnetic moments of its neighboring semi-magic nuclei, $^{41}$Sc and $^{39}$Ca, which occur near the doubly magic $^{40}$Ca according to the additivity rule. $^{41}$Sc and $^{39}$Ca are doubly magic plus or minus one nucleon nuclei and their magnetic moments are supposed to be close to the single particle values. However, it has been shown that sizable cross shell excitations are required to reproduce their magnetic moments \cite{min90, klo19}. In addition to shell model calculations, recently developed valence-space in-medium similarity renormalization group calculations for multiple major-oscillator shells were also applied for the magnetic moment of $^{40}$Sc to benchmark such calculations for an odd-odd nucleus requiring cross shell excitations.

\section{Hyperfine Interaction}
\label{sec:hf-interaction}

The shift of an atomic energy level due to the hyperfine (hf) interaction relative to an atomic fine-structure level is given by
%
	\begin{align}
	\frac{\Delta E}{h} &= \alpha A^{\mathrm{hf}} + \beta B^{\mathrm{hf}},\nonumber\\
	\alpha &=  \frac{K}{2},\nonumber\\
	\beta &= \frac{3K(K+1)-4I(I+1)J(J+1)}{8I(2I-1)J(2J-1)},\nonumber\\
	K & = F(F+1) - I(I+1) - J(J+1).
	\label{eq:hyperfine_shift}
	\end{align}
%
Here $I$ and $J$ are the total angular momenta or spins of the nucleus and electrons, respectively, and $F$ is the quantum number associated with the total angular momentum of the atom defined by $\mathbf{F}=\mathbf{I}+\mathbf{J}$. The $A^{\mathrm{hf}}$ and $B^{\mathrm{hf}}$ are the hyperfine coupling constants defined as
%
	\begin{align}
		A^{\mathrm{hf}} &= \frac{\mu B(0)}{IJ},\label{eq:hyperfine_A}\\
		B^{\mathrm{hf}} &= eQ \left\langle \frac{\partial^{2}V_{e}}{\partial z^{2}} \right\rangle,\label{eq:hyperfine_B}
	\end{align}
%
where $B(0)$ is the magnetic field generated by the atomic electrons at the center of the nucleus, $e$ is the electric unit charge, $Q$ is the spectroscopic nuclear electric-quadrupole moment, and $\langle \partial^{2}V_{e}/\partial z^{2} \rangle$ is the electric field gradient produced by the atomic electrons at the center of the nucleus.
The magnetic field and the electric field gradient are isotope independent assuming a point-like nucleus (the hyperfine anomalies~\cite{boh50} are neglected here). 
According to Eqs.~(\ref{eq:hyperfine_A}) and (\ref{eq:hyperfine_B}), unknown nuclear moments may be deduced from the measured hyperfine coupling constants using a reference nucleus of the same element, whose hyperfine coupling constants for the same electronic level and nuclear moments are known.
A simple ratio of hyperfine coupling constants derives nuclear moments as
%
	\begin{align}
		\mu &= \mu_{\mathrm{R}}\frac{A^{\mathrm{hf}}}{A_{\mathrm{R}}^{\mathrm{hf}}}\frac{I}{I_{\mathrm{R}}},\label{eq:hyperfine_A_2}\\ 
		Q &= Q_{\mathrm{R}}\frac{B^{\mathrm{hf}}}{B_{\mathrm{R}}^{\mathrm{hf}}},\label{eq:hyperfine_B_2}
	\end{align}
%
where the subscript R indicates a reference nucleus. If a suitable reference isotope does not exist, the magnetic field and electric field gradient may be obtained by theoretical calculations. In the present study $^{45}$Sc was used as the reference isotope. 

The transition frequency $\nu$ of an electronic transition between a pair of hyperfine levels can be expressed using five parameters as
%
\begin{equation}
\label{eq:trans_freq}
  \nu = \nu_{0} + (\alpha_{u}A^{\mathrm{hf}}_{u} + \beta_{u}B^{\mathrm{hf}}_{u}) - (\alpha_{l}A^{\mathrm{hf}}_{l} + \beta_{l}B^{\mathrm{hf}}_{l}),
\end{equation}
%
where $\nu_{0}$ is the center-of-gravity (COG) frequency, and the $u$ and $l$ subscripts refer to the upper and lower hyperfine states of the transition, respectively. The $\alpha$ and $\beta$ are constants determined by the nuclear and atomic spins of the specific transition given by Eq.~(\ref{eq:hyperfine_shift}).

\section{Experiment}
\label{sec:experiment}
The radioactive $^{40}$Sc ion beam was produced by the one-proton pickup and one-neutron removal reaction of a primary \textsuperscript{40}Ca beam on a natural Be target. The $^{40}$Ca beam was accelerated to 140 MeV/A in the coupled cyclotrons at the National Superconducting Cyclotron Laboratory at Michigan State University. The produced fast \textsuperscript{40}Sc ion beam was separated in the A1900 fragment separator \cite{mor03} from other reaction products, thermalized in a gas cell \cite{sum20} filled with a helium buffer gas, and extracted by RF and DC electric fields as singly-charged ions at an energy of 30 keV. 

The low-energy $^{40}$Sc ion beam was then transported to the beam cooler and laser spectroscopy (BECOLA) facility \cite{min13, ros14}, where the ion beam was injected into a radio frequency quadrupole (RFQ) cooler-buncher \cite{bar17} filled with a helium-buffer gas. Buffer gas pressures in the RFQ were $\sim$100 mTorr and $\sim$1 mTorr in the cooling and bunching sections, respectively. The ions were collected for 300 ms and then released in bunches with a typical temporal width of 1 $\mu$s (full-width at half-maximum) at an approximate energy of 29,850 eV and transported to the collinear laser spectroscopy (CLS) beam line. By selecting photons only within the time window of beam bunches \cite{ros14}, the photon background originating from scattered laser light can be dramatically reduced \cite{cam02, nie02}. In the present measurement a background suppression factor was 3 $\times$ 10$^5$.

The ions were then collinearly overlapped with laser light for laser-resonant fluorescence measurement. Fluorescence was detected by a photon detection unit consisting of a mirror based light collector \cite{maa20}, which eliminates background and focuses resonant fluorescence onto a photomultiplier tube. A scanning voltage was applied to the photon detection unit, which is electronically isolated from the ground potential to vary the ion beam velocity and to scan the Doppler-shifted laser frequency across the hyperfine spectrum. The voltage scanning range is typically \SI{-2}{\kilo\volt} to \SI{2}{\kilo\volt}. The induced fluorescence was then detected by the photon detection unit as a function of the applied scanning voltage.

Typical rates  of $^{40}$Sc and $^{41}$Sc at the entrance to the RFQ were 15\,$\times$\,10$^3$  and 20\,$\times$\,10$^3$ ions/s, respectively. There was a large contamination of $^{40}$Ar ions ($> $\,$10^7$/s) in the $^{40}$Sc beam, due to the trace amount argon component in the helium buffer gas of the gas stopping cell. The RFQ can accept typically 10$^6$ ions for trapping and bunching. When more ions are injected, some of ions are lost and measured hyperfine spectra become distorted due to the excess space charge in the trap \cite{bar17}. In order to avoid this overfilling, $^{40}$Sc hyperfine spectra were measured with faster bunching cycle (20 ms) at the cost of the background suppression. The total data accumulation time for $^{40}$Sc and $^{41}$Sc were 70 and 30 hours, respectively. The longer accumulation time for $^{40}$Sc is mainly due to the $^{40}$Ar contamination.

The hyperfine spectrum of $3d4s\,^3\text{D}_2\leftrightarrow3d4p\,^3\text{F}^\circ_3$ transition in Sc II at 363 nm was measured. Laser light at 726 nm was produced by a Sirah Matisse TS Ti:sapphire ring laser, which was then frequency doubled to 363 nm using a SpectraPhysics Wavetrain. The laser frequency of the Ti:sapphire laser was measured and stabilized by a HighFinesse WSU-30 wavelength meter, calibrated every minute by a frequency stabilized helium-neon laser. The 363 nm laser light was collimated to a beam diameter of approximately \SI{1}{\milli\meter}. The laser light was aligned to go through the centered of the photon-detection region using two 2.5-m apart \SI{3}{\milli\meter} apertures in the BECOLA beamline. Three photon detection units were placed in series along the ion beam direction to detect the resonant fluorescence. 

Every few hours throughout the data accumulation, the radioactive Sc beam was stopped and a stable $^{45}$Sc beam was introduced into the CLS beam line for about an hour for calibration purposes. The $^{45}$Sc ion beam was produced using a Penning Ionization Gauge ion source \cite{ryd15} installed upstream of the RFQ. Fluorescence spectra of $^{45}$Sc were measured as a reference to determine the resonance line shape and to monitor the time-dependent centroid shift due mainly to the voltage shift caused by the temperature variation over a week-long running time. The number of $^{45}$Sc ions in a bunch was limited to about 10$^5$ to keep a similar space charge condition in the trap as to radioactive beam measurements, but the data collection was done with a higher beam-bunch repetition rate of 100 Hz to efficiently collect calibration data. 

\begin{table*}[!htbp]
    \caption{Hyperfine coupling constants for the $^3\text{D}_2$ $\leftrightarrow$ $^3\text{F}^\circ_3$ electronic transition in \textsuperscript{40,41,45}Sc extracted from the fit of the hyperfine spectra. For \textsuperscript{40,41}Sc the $^3\text{D}_1$ $\leftrightarrow$ $^3\text{P}_0$ transition was also measured, and coupling constants for the $^3\text{D}_1$ state are also summarized. The low statistics of the \textsuperscript{40}Sc data gives possible uncertainties in the fitting procedure. These fitting uncertainties were estimated from the variation of several different fitting procedures and were included in the uncertainties shown for \textsuperscript{40}Sc.}
    \begin{tabularx}{0.999 \textwidth}{X>{\centering\arraybackslash}X>{\centering\arraybackslash}X>{\centering\arraybackslash}X>{\centering\arraybackslash}X>{\centering\arraybackslash}X>{\centering\arraybackslash}X}
    \hline
	Isotope & \multicolumn{3}{c}{$A^\text{hf}$ (MHz)} &  \multicolumn{3}{c}{$B^\text{hf}$ (MHz)}  \\
	\cmidrule(){1-1}\cmidrule(r){2-4}\cmidrule(l){5-7}
        & $^3\text{D}_1$ & $^3\text{D}_2$ & $^3\text{F}^\circ_3$ & $^3\text{D}_1$ & $^3\text{D}_2$ & $^3\text{F}^\circ_3$ \\
         \cmidrule(r){2-4}\cmidrule(l){5-7}
        \textsuperscript{40}Sc(4$^-$) & & +520.1(39)(19) & +210.7(16)(08)\footnote{The ratio to $A^{\rm hf}(^3{\rm D}_2)$ was fixed relative to that of $^{45}$Sc.} &  & +58(52)(39)         & +106(95)(66)\footnote{The ratio to $B^{\rm hf}(^3{\rm D}_2)$ was fixed relative to that of $^{45}$Sc.}\\
         \textsuperscript{41}Sc(7/2$^-$) & -549.4(16)(1) & +579.7(12)(01) & +234.81(76)(04) & -16(10)(0) & -25(14)(0)        & -26(17)(0)              \\
        \textsuperscript{45}Sc(7/2$^-$) & -479.58(3)(3) & +507.54(2)(4) & +205.57(17)(12)  & -11.7(1)(0) & -32.63(21)(15)     & -59.27(24)(5)         \\
        \hline
    \end{tabularx}
    \label{tab:hyperfine_results}
\end{table*}

\begin{figure}[t]
    \centering
          \includegraphics[width=\linewidth]{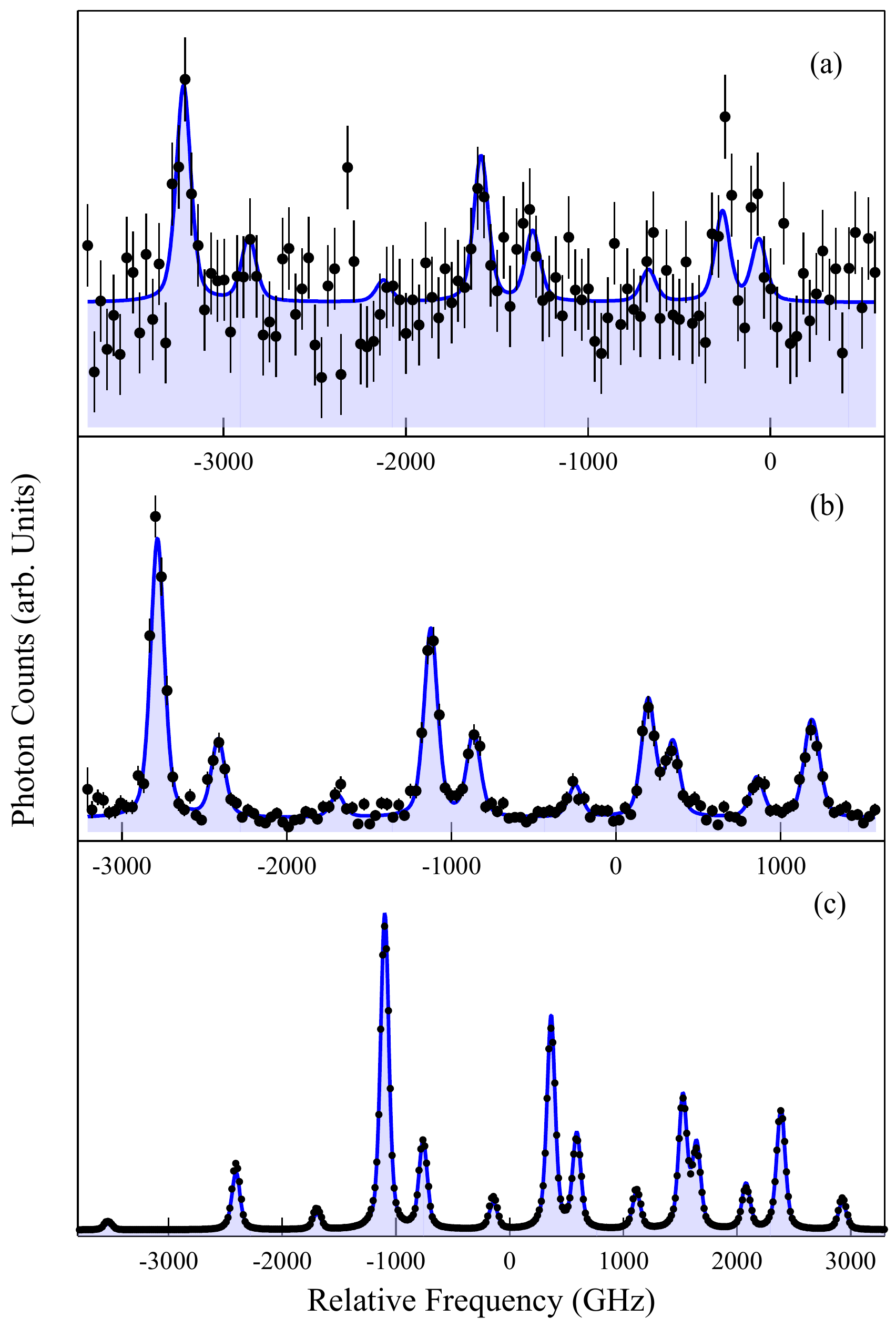}
    \caption{Obtained hyperfine spectra of $3d4s\,^3\text{D}_2\leftrightarrow3d4p\,^3\text{F}^\circ_3$ transition in Sc II for (a) \textsuperscript{40}Sc , (b) \textsuperscript{41}Sc, and (c) \textsuperscript{45}Sc. The horizontal-axis are relative to the center of gravity resonance frequency of the \textsuperscript{45}Sc hyperfine spectrum. The vertical bars are statistical uncertainties. For $^{45}$Sc the statistical uncertainties are within the dots and not shown in the figure. The solid blue lines and shades are results of the fits. For $^{40}$Sc and $^{41}$Sc not all allowed transitions were observed due to the sensitivity of the detection unit.}
    \label{fig:hyperfine_spectra}
\end{figure}

\section{Experimental Results}
\label{sec:results}

\subsection{Hyperfine spectra}
The obtained hyperfine spectra for the $3d4s\,^3\text{D}_2\leftrightarrow3d4p\,^3\text{F}^\circ_3$ transition are shown in Fig.\;\ref{fig:hyperfine_spectra} for $^{40}$Sc, $^{41}$Sc, and $^{45}$Sc. The solid dots are the data and the solid line is a result of a line shape fit. \textsuperscript{40}Sc, \textsuperscript{41}Sc and \textsuperscript{45}Sc have nuclear spins of 4, 7/2 and 7/2 respectively and have 15 allowed hyperfine transitions, which were not all observed for $^{40}$Sc and $^{41}$Sc, due to the achievable signal to noise ratio in a limited beam time given the sensitivity of the detection system. Especially for $^{40}$Sc, a large amount of $^{40}$Ar contamination in the beam prevented the background suppression by the bunched beam CLS technique, due to the overfilling of the RFQ ion trap. 

The hyperfine spectra were fit using a PseudoVoigt line shape with all peaks having a common width and Lorentzian-Gaussian fraction. The relative peak amplitudes for all spectra were fixed to ratios among transition probabilities of hyperfine transitions determined by the quantum numbers ($I$, $J$, and $F$) of the lower and upper levels, and the overall amplitude was determined by the fit. The individual peak frequency can be expressed using five parameters as in Eq. (\ref{eq:trans_freq}), namely $A^\text{hf}$ and $B^\text{hf}$ coefficients for lower and upper levels of the transition and the COG frequency, all of which were allowed to vary freely in the fitting procedure.

Due to the poor signal to noise ratio of the \textsuperscript{40}Sc data, the peak width and Lorentzian-Gaussian fraction determined by the hyperfine spectrum of \textsuperscript{41}Sc, which has greater signal-to-noise ratio, were used to constrain the fit of the \textsuperscript{40}Sc spectrum. Furthermore, the hyperfine coupling constants were constrained to match the ratio of coupling constants of our precise measurements of stable \textsuperscript{45}Sc as $A^{\rm hf}_l$/$A^{\rm hf}_u$ = 2.4689(5) and $B^{\rm hf}_l$/$B^{\rm hf}_u$ = 0.55(2). The uncertainties of the ratios were taken into account as systematic uncertainties of $^{40}$Sc hyperfine coupling constants, which are small compared to the statistical uncertainties of the $^{40}$Sc results. With these constraints and known nuclear spin of 4, the fit result was uniquely obtained from the pattern of the observed hf spectrum, though not all allowed transitions were detected. 

\begin{table*}[!htbp]
\caption{Ground state electromagnetic moments of $^{40, 41, 45}$Sc. The \textsuperscript{41}Sc electromagnetic moments were deduced from the weighted average of the moments calculated using the upper and lower level hyperfine coupling constants of both the $^3\text{D}_2\leftrightarrow ^3\text{F}^\circ_3$ and $^3\text{D}_1\leftrightarrow ^3\text{P}_0$ transitions. Previously measured values of the \textsuperscript{41, 45}Sc electromagnetic moments are also included in the table. The signs are not assigned experimentally, where it is given in a parenthesis.}
    \begin{tabularx}{0.99 \textwidth}{>{\centering\arraybackslash}X>{\centering\arraybackslash}X>{\centering\arraybackslash}X>{\centering\arraybackslash}X>{\centering\arraybackslash}X>{\centering\arraybackslash}X}
    \hline
    Isotope & $I^\pi$ & \multicolumn{2}{c}{$\mu\;(\mu_N$)} & \multicolumn{2}{c}{$Q\;(e^2 {\rm fm^2)}$} \\
     \cmidrule(){1-2}\cmidrule(r){3-4}\cmidrule(l){5-6}
    && this work & Lit. & this work & Lit. \\
    \cmidrule(r){3-4}\cmidrule(l){5-6}
  	$^{40}$Sc & 4$^-$ & +5.57(4)(2) &  & +42(38)(28) & \\
        $^{41}$Sc & 7/2$^-$ & +5.4376(80)(06) & (+)5.431(2) \cite{min90} & -18.5(71)(01) & (-)15.6(3) \cite{min02}\\
        $^{45}$Sc & 7/2$^-$ & & +4.7563(5)\footnote{This value was re-evaluated based on \cite{lut69} in the present work. See text for detail.} & & -23.6(2)\footnote{This value was re-evaluated in \cite{min02}.} \\
        \hline
    \end{tabularx}
    \label{tab:moments}
\end{table*}

\subsection{Corrections}
The COG frequency of the $^3\text{D}_2\,\leftrightarrow\,^3\text{F}^\circ_3$ transition was determined to be 825,470,305(2)(20) MHz in a separate offline measurement using collinear and anti-collinear laser spectroscopy technique on stable $^{45}$Sc. The first parentheses of the COG frequency is for the statistical uncertainty and the second is for the systematic uncertainty due to the laser frequency measurement using the WSU-30 wavelength meter. The COG frequency was used to calibrate the Sc ion beam energy in the online $^{40}$Sc hyperfine spectrum measurements. The $^{45}$Sc calibration spectra were used to deduce the ion beam energy as described in \cite{koenig_21}, where the 20 MHz systematic uncertainty in the COG is mostly canceled. The ion beam energy was determined with less than 0.3 eV uncertainty, which contributes to the systematic uncertainties of the stable hyperfine coupling constants by less than 10 ppm.

A correction for the nonlinearity of scanning voltage was applied to the hyperfine spectra due to a penetration of the static electromagnetic field applied to the detection region for the Doppler tuning. The nonlinear behavior in the scanning voltage would stretch or shrink the hyperfine spectrum observed, which would then affect the deduced hyperfine coupling constants. This correction was estimated by comparing the isotope shift between $^{40}$Ca and $^{44}$Ca measured at BECOLA to the well known literature values. A correction of 550(50) ppm was estimated and applied to the present hyperfine coupling constants. The uncertainty  is included in the systematic uncertainties of the hyperfine coupling constants. The voltmeter used to measure the scanning voltage was rated to have an approximate uncertainty of 80 ppm, which is also included in the systematic uncertainties. Both contributions to the total systematic uncertainty are negligibly small compared to the statistical and other systematic uncertainties of $^{40}$Sc.

Determined hyperfine coupling constants are summarized in Table \ref{tab:hyperfine_results}. Signs are explicitly given for all experimental numbers hereafter. The statistical and systematic uncertainties are shown in the first and second parenthesis, respectively. For $^{40}$Sc an additional systematic contribution is considered resulting from the correlation between $A^{\rm hf}$ and $B^{\rm hf}$ coupling constants in the fitting procedure. Especially the $B^{\rm hf}$ has large uncertainty that affects the determination of $A^{\rm hf}$. To estimate the effect of the uncertainty of $B^{\rm hf}$ on $A^{\rm hf}$, we varied and fixed $B^{\rm hf}$ in the fitting procedure, and take the variation in $A^{\rm hf}$ as the systematic uncertainty of $A^{\rm hf}$.

\subsection{Magnetic dipole moment}
The magnetic-dipole hyperfine coupling constant of the $3d4s$\;$^3{\rm D}_2$ state in $^{40}$Sc was obtained as 
%
	\begin{equation}	
		A^{\rm hf}(^3{\rm D}_2) = +520.1 (39)(18) \,\rm{MHz}.\nonumber\label{eq:hyperfine_A_exp}
	\end{equation}
%
The magnetic moment of $^{40}$Sc can be deduced from Eq.~(\ref{eq:hyperfine_A_2}) together with the $A^{\rm hf}$($^3{\rm D}_2$) and the magnetic moment of $^{45}$Sc as
%
	\begin{equation}
		\mu(^{40}\mathrm{Sc}) = \mu(^{45}\mathrm{Sc})\frac{A^{\mathrm{hf}}(^{40}\mathrm{Sc})}{A^{\mathrm{hf}}(^{45}\mathrm{Sc})}\frac{I(^{40}\mathrm{Sc})}{I(^{45}\mathrm{Sc})}.
		\label{eq:hyperfine_a_calc}
	\end{equation}
%
Here it is noted that deviation from a point-like nucleus, the hyperfine anomaly \cite{boh50}, and its effect on hyperfine coupling constants is not known for any of the atomic states in Sc isotopes \cite{per13}. The hyperfine anomaly is typically smaller than the relative uncertainty of $\sim$~1\% in the present result of magnetic moment, and is therefore neglected in the present analysis. 
The magnetic moment of reference $^{45}$Sc was evaluated from the ratio of NMR frequencies between $^{45}$Sc (in a solution of ScCl$_3$ in acidified deuterium water) and deuterium $\nu(^{45}\mathrm{Sc})/\nu(D) = 1.5824534 (6)$ \cite{lut69}. Using values of the ratio between NMR frequencies of hydrogen in H$_2$O and deuterium in D$_2$O $\nu(D)/\nu(p) = 0.153506083(60)$ \cite{sma51}, the diamagnetic shielding constant for Sc$^{3+}$ in ScCl$_3$ $\sigma(^{45}\mathrm{Sc}) = 0.00156(10)$ \cite{joh83} and proton in water $\sigma(p) = 25.687(15) \times 10^{-6}$ \cite{moh16}, and magnetic moment of proton $\mu(p) = +2.792847 337(29) \mu_N$ \cite{moh16}, the magnetic moment of $^{45}$Sc can be obtained \cite{ant05} as 
%
\begin{align}
	\mu(^{45}\mathrm{Sc}) &= \frac{\nu(^{45}\mathrm{Sc})}{\nu(p)}\frac{1-\sigma(p)}{1-\sigma(^{45}\mathrm{Sc})}\frac{I(^{45}\mathrm{Sc})}{I(p)}\mu(p)\nonumber\\
	& = +4.75631(48)\;\mu_N.\nonumber
\end{align}
%
Together with our present measurement of $A^{\mathrm{hf}}(^{45}\mathrm{Sc}) = +507.54(2)(4)\,\mathrm{MHz}$, the ground state magnetic moment of $^{40}$Sc ($I^\pi = 4^-$) was deduced from Eq. (\ref{eq:hyperfine_a_calc}) to be 
	\begin{equation}
		\mu(^{40}\mathrm{Sc}) = +5.57 (4) (2)\,\mu_N,\nonumber
	\end{equation}
which is also summarized in Table~\ref{tab:moments}. Here the correlation between the variation of $A^{\rm hf}$ due to possible variation of $B^{\rm hf}$ was taken into account and propagated in the systematic uncertainty of the present $\mu(^{\rm 40}$Sc).

\subsection{Electric quadrupole moment}
The electric-quadrupole hyperfine coupling constant of $^{40}$Sc was obtained as 
%
	\begin{equation}	
		B^{\rm hf}(^3{\rm D}_2) = +58 (52) (39) \,\rm{MHz}.\nonumber
		\label{eq:hyperfine_B_exp}
	\end{equation}
%
The quadrupole moment of $^{40}$Sc can be deduced from Eq.~(\ref{eq:hyperfine_B_2}) together with the present $B^{\rm hf}$($^3{\rm D}_2$) for $^{40}$Sc and $^{45}$Sc, and the quadrupole moment of $^{45}$Sc as
%
	\begin{equation}
		Q(^{40}\mathrm{Sc}) = Q(^{45}\mathrm{Sc})\frac{B^{\mathrm{hf}}(^{40}\mathrm{Sc})}{B^{\mathrm{hf}}(^{45}\mathrm{Sc})}.
		\label{eq:hyperfine_b_calc}
	\end{equation}
%
Using values of our measurement of $B^{\mathrm{hf}}(^{45}\mathrm{Sc}) = -32.63 (21)(15)\,\mathrm{MHz}$, $Q(^{45}\mathrm{Sc}) = -23.6 (2)\,e~{\rm fm^2}$~\cite{min02}, the ground state quadrupole moment of $^{40}$Sc ($I^\pi = 4^-$) was deduced to be 
%
	\begin{equation}
		Q(^{40}\mathrm{Sc}) = +42 (38) (28)\,e\;{\rm fm^2}.\nonumber
	\end{equation}
%
No further meaningful discussion can be made for the quadrupole moment of $^{40}$Sc due to the large uncertainty. All values are summarized in Table~\ref{tab:moments}.

\begin{table*}[t]
    \caption{Shell model and multi-shell IMSRG calculations for magnetic moment of the 4$^-$ ground state in $^{40}$Sc and $^{40}$K, together with their experimental values (Exp). Magnetic moments of the $A$ = 39 and 41 systems are also listed to obtain values based on the additivity rule. All values are given in a unit of $\mu_N$. For the shell model, sp, full and emp are for the single particle, full and corrected empirical calculations, respectively. See text for more details.}
    \begin{tabularx}{0.99 \textwidth}{>{\raggedright\arraybackslash}X>{\centering\arraybackslash}X>{\centering\arraybackslash}X>{\centering\arraybackslash}X>{\centering\arraybackslash}X>{\centering\arraybackslash}X>{\centering\arraybackslash}X>{\centering\arraybackslash}X>{\centering\arraybackslash}X}
    \hline
    & \multicolumn{3}{c}{Shell Model} & \multicolumn{3}{c}{IMSRG} & Exp & Ref \\
   \cmidrule(r){2-4}\cmidrule(l){5-7}
    & sp & full & emp & EM\footnote{1.8/2.0(EM)} & NNLO\footnote{NNLOgo(394)} & N3LO\footnote{N3LO$_{lnl}$} & & \\
    \hline
    $\mu(^{40}$K) & & -1.707\footnote{Calculated with free nucleon g-factors.} &  & -1.337 & -1.096 & -1.327 & -1.2981(0) & \cite{sah74} \\
    $\mu(^{40}$Sc) & & 5.936\footnote{Calculated with free nucleon g-factors.} & & 5.585 & 5.400 & 5.583 & +5.570(40)(20) & Present \\
    $\mu_{\rm IS}$ & & 2.115 & & 2.124 & 2.152 & 2.128 & +2.136(20) &\\
    $\left<s_3\right>$ & & 0.301 & & 0.326 & 0.399 & 0.337 & +0.358(53) &\\
    $\mu_{\rm IV}$ & & 3.822 & & 3.461 & 3.248 & 3.455 & +3.434(20) &\\
    \hline
    $\mu(^{39}$K) & 0.124 & -0.469 &  & -0.019 & 0.025 & -0.034 & -0.3915(0) & \cite{sah74} \\
    $\mu(^{39}$Ca) & 1.148  & 0.930 &  & 1.337 & 1.326 & 1.369 & (+)1.0217(1) & \cite{min76} \\
    $\mu(^{41}$Ca) & -1.913 & -2.021 &  & -1.308 & -1.236 & -1.320 & -1.5947(0) & \cite{bru62} \\
    $\mu(^{41}$Sc) & 5.793  & 5.813 &  & 5.199 & 5.113 & 5.203 & (+)5.431(2) & \cite{min90} \\
    \hline
    &\\
    \hline
    & \multicolumn{7}{c}{using additivity rule} &\\
    \hline
    $\mu(^{40}$K) & -1.683 & -1.598 & -1.273 & -1.206 & -1.116 & -1.224 & -1.2492(0) & \\      
    $\mu(^{40}$Sc) & 5.909 & 5.811 & 5.538 & 5.476 & 5.382 & 5.487 & (+)5.510(2) \\
    $\mu_{\rm IS}$ & 2.113 & 2.11 & 2.133 & 2.130 & 2.133 & 2.131 & (+)2.131(1) &\\
    $\left<s_3\right>$ & 0.297 & 0.28 & 0.349 & 0.342 & 0.350 & 0.345 & (+)0.344(2) &\\
    $\mu_{\rm IV}$ & 3.796 & 3.70 & 3.406 & 3.336 & 3.249 & 3.356 & (+)3.380(1) &\\
    \hline
    \end{tabularx}
    \label{tab:summary}
\end{table*}

\section{Discussion}
\label{sec:discussion}

\subsection{Shell model calculations}
The magnetic moment operator can be expressed as 
%
\begin{equation}
\bm{\mu} = g_s\bm{s} + g_l\bm{l},
\end{equation}
%
where $g_s$ and $g_l$ are the spin and the orbital $g$ factors, respectively. The free nucleon $g$ factors ($g_l^\mathrm{p} = 1$, $g_l^\mathrm{n} = 0$, $g_s^\mathrm{p} = 5.586$ and $g_s^\mathrm{n} = -3.826$) were used for the calculation.  For the magnetic moment of $^{40}$Sc we used a shell-model space for one proton in the full $fp$ shell $(0f_{7/2}, 0f_{5/2}, 1p_{3/2}, 1p_{1/2})$, and one neutron hole in the full $sd$ shell $(0d_{3/2}, 1s_{1/2}$, $0d_{5/2})$ with the $sdpf$-$wb$ effective shell-model interactions \cite{sdpfw}. The $I^\pi$ = 4$^-$ ground state magnetic moment with the full wavefunction was obtained as
%
\begin{equation}
\mu(^{40}{\rm Sc})_{\rm full} = 5.936\;\mu_N.\nonumber
\end{equation} 
%
This model space and interaction were also used recently with regard to high-spin states in $^{40}$Sc \cite{ga20}. The 4$^{-}$ wavefunction of the ground state is dominated (97.97\%) by the canonical $(\pi$:\,$0f_{7/2})(\nu$:\,$0d_{3/2}^{-1})$ configuration. For this almost pure configuration we can use the additivity rule to obtain the magnetic moments of $^{40}$Sc. According to Eq.\,(\ref{eq:additivity}) we obtain
%
\begin{equation}
\mu(^{40}{\rm Sc}) = \frac{32}{35}\;\mu_p(0f_{7/2}) + \frac{8}{15}\;\mu_n(0d_{3/2}^{-1}).\nonumber
\end{equation}
%
Using the single particle magnetic moments of $\mu_p(0f_{7/2})$ and $\mu_n(0d_{3/2}^{-1})$ obtained from Eq.\;(\ref{eq:spmagneticmom}), this pure configuration has a free-nucleon single-particle (sp) magnetic moment (in a unit of $\mu_N$) of
%
\begin{equation}
\mu(^{40}{\rm Sc})_{\rm sp} = \frac{32}{35}\;5.793 + \frac{8}{15}\;1.148 = 5.909.\nonumber
\end{equation}
%
Alternatively we can evaluate to some approximation the pure configuration using known empirical ground state magnetic moments of $^{41}$Sc(7/2$^-$) \cite{min90} and $^{39}$Ca(3/2$^+$) \cite{min76} for the $\mu_p(0f_{7/2})$ and $\mu_n(0d_{3/2}^{-1})$, respectively. This would take into account the components for the wavefunction that go beyond one-particle one-hole configuration, and mesonic exchange corrections to the free nucleon g-factors. In a unit of $\mu_N$ this magnetic moment (pure) is given
%
\begin{equation}
\mu (^{40}{\rm Sc})_{\rm pure} = \frac{32}{35}\;5.431(2) +  \frac{8}{15}\;1.0217(1) = 5.510(2),\nonumber
\end{equation}
%
where numbers in the parentheses are experimental uncertainties of magnetic moments, which are small compared to the present result and will be ignored hereafter.  

\begin{figure}[t]
        \includegraphics[width=\linewidth]{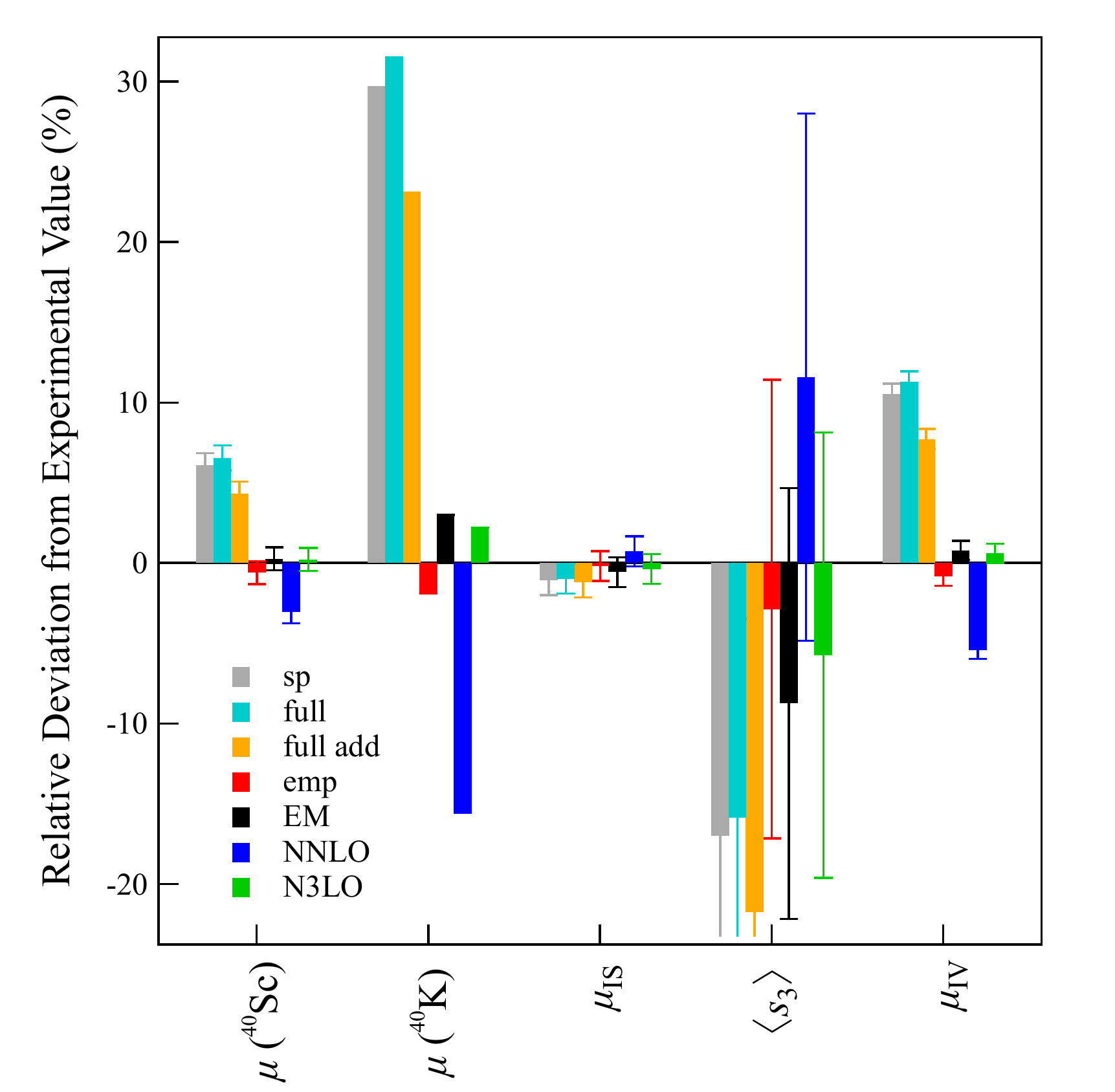}
    \caption{Relative deviations of theoretical magnetic moments from experimental values are shown. The sp, full, emp are for the single particle, full and empirical shell model calculations, respectively (see text for more detail). The EM, NNLO and N3LO are for IMSRG calculations with 1.8/2.0(EM), NNLOgo(394) and N3LO$_{lnl}$ interactions, respectively. Uncertainties are also included, which are dominated by the experimental uncertainty in the present $\mu$($^{40}$Sc).}
    \label{fig:result}
\end{figure}

The difference of the calculated magnetic moment of the full particle-hole configuration and the simple single particle $(0f_{7/2})(0d_{3/2}^{-1})$ configuration with free nucleon g-factors is $\mu_{\rm full} - \mu_{\rm sp} = 0.028\;\mu_N$. Adding this small difference to $\mu (^{40}{\rm Sc})_{\rm pure}$, we obtain a corrected magnetic moment as 
%
\begin{equation}
\mu (^{40}{\rm Sc})_{\rm emp} = 5.538\;\mu_N.\nonumber
\end{equation}
%
This is an empirical shell-model (emp) value obtained within the additivity rule corrected for the configuration mixing and meson-exchange effect, which is in good agreement with the present experimental value of 5.57(4)(2)$\mu_N$. The results are summarized in Table\;\ref{tab:summary} and Figure \ref{fig:result} together with results of similar analysis for the mass $A$ = 40 mirror nucleus $^{40}$K(4$^-$), whose ground state magnetic moment is known \cite{sah74}. Here magnetic moments of $^{39}$K(3/2$^+$) \cite{sah74} and $^{41}$Ca(7/2$^-$) \cite{bru62} were used as the pure configuration for the $\mu_p(0d_{3/2}^{-1})$ and $\mu_n(0f_{7/2})$, respectively, to extract the $\mu(^{40}{\rm K})_{\rm emp}$ from the additivity rule. The magnetic moment calculated using the additivity rule with the correction $\mu_{\rm full} - \mu_{\rm sp} = -0.024\;\mu_N$ agrees well with the experimental value. For both magnetic moments of $^{40}$Sc and $^{40}$K, the $\mu_{\rm sp}$ and $\mu_{\rm full}$ deviate significantly from the experimental value, indicating departure from the canonical configuration and needs for mesonic effects. It is noted that calculations for $^{40}$K show larger deviation from the experimental magnetic moment than those for $^{40}$Sc.

The magnetic moments for the $A$ = 39 obtained with the empirical $  sd  $ shell effective g factors \cite{usd} and for the $A$ = 41 obtained with the empirical $  fp  $ shell effective g factors \cite{hon04} are also summarized in Table \ref{tab:summary} as "full". The magnetic moments for the $A$ = 40 deduced using these values using the additivity rule are also listed in Table~\ref{tab:summary} and shown in Figure \ref{fig:result} as "full add". The magnetic moments obtained with the effective g-factors are slightly closer to the experimental values than those with the free g-factors (sp) and the full calculations, but still deviates significantly from the experimental values. It can be  regarded that the difference with experiment as being due to excitations of nucleons from $  sd  $ to $  fp  $ shell in nuclei near $^{40}$Ca \cite{min90,klo19} that are in excess of those present in the middle of the $  sd  $ and $  fp  $ shells. Our analysis shows that these excess excitations do not change the additivity relationships expected from the simple particle-hole wavefunctions as can be seen from the good agreement of the empirical shell-model calculations with the experimental value of magnetic moments.

\subsection{ab-initio IMSRG calculation}
In addition, we provide ab initio magnetic moments from the valence-space formulation of the ab initio in-medium similarity renormalization group (VS-IMSRG) \cite{Herg16PR,Stro19ARNPS} based on input two- and three-nucleon forces from chiral EFT~\cite{Epel09RMP,Mach11PR}. Ab initio calculations of electromagnetic properties have progressed to regions near the $sd$ shell~\cite{klo19,Hend18E2,Heil20OE2,Ciem20OE2,Heyl21Al}, where notable, but largely understood, discrepancies have emerged compared with data. Using the Magnus approach~\cite{Morr15Magnus}, we generate an approximate unitary transformation to decouple a given core and valence-space Hamiltonian from the full-space Hamiltonian, where operators are truncated at the two-body level (IMSRG(2)), and 3N forces between valence nucleons is captured via ensemble normal ordering~\cite{Stro17ENO}. The same transformation is used to decouple effective valence-space $M1$ operators consistent with the Hamiltonian~\cite{Parz17Trans}, in principle eliminating the need for quenching typically seen in phenomenological approaches. We note, however, that effects of two-body currents are expected to be non-negligible but are not currently implemented.

We start from a harmonic-oscillator basis of 15 major shells and use three different NN+3N interactions derived from chiral EFT. First the 1.8/2.0(EM)~\cite{Hebe11fits,Simo17SatFinNuc} and N3LO$_{lnl}$ are known to well reproduce ground-state energies globally to the tin region~\cite{Morr18Tin,Stro21Drip}, while underpredicting charge radii~\cite{Groo20Cu}. The NNLOgo(394) interaction includes explicit delta isobar degrees of freedom~\cite{Jian20N2LOgo} and generally exhibits improved radii properties without sacrificing energies~\cite{Kosz21Krad,Moug21Sn}. Since the physics of $^{40}$Sc is expected to span the $sd$ and $pf$ shells, we use the newly formulated multishell approach~\cite{Miya20lMS} to decouple a ${}$ valence-space Hamiltonian and effective $M1$ operator above a $^{28}$Si core, in an increased $E_\mathrm{3max} = 24$ truncation on storage of 3N matrix elements~\cite{Miya21Heavy}. Finally, energies and moments are obtained from the KSHELL shell model code~\cite{KSHELL}, where only free $g$-factors are used. Though effects of two-body current are not included in the calculations, the overall agreement with experimental magnetic moments of $^{40}$K and $^{40}$Sc are good but the NNLOgo(394) interaction, which show larger deviation from experimental value than other two interactions, especially for $^{40}$K.

\subsection{Isoscalar and isovector magnetic moments}
Examination of only the contribution from the isoscalar (IS) and isovector (IV) parts of the magnetic moment can also provide insight into shell structure and configuration mixing. The IS and IV magnetic moment can be obtained from mirror magnetic moments as
%
\begin{align}
\mu_{\rm IS} &= \frac{1}{2}\left\{\mu(T_3 = +T) + \mu(T_3 = -T)\right\}\nonumber\\
\mu_{\rm IV} &= \frac{1}{2}\left\{\mu(T_3 = +T) - \mu(T_3 = -T)\right\}
\end{align}
%
with the isospin $T_3$ = +1/2 for protons. Assuming good isopspin symmetry and ignores the isoscalar mesonic exchange currents, the isoscalar spin expectation value $\left<s_3\right>$ \cite{sac46, sug69}, which is the contribution of nuclear spins to the magnetic moment, can also be evaluated from the $\mu_{\rm IS}$ as
$\left<s_3\right> = (\mu_{\rm IS} - I/2)/(g^s_p + g^s_n - g_p^l)$, where $g^s_p$, $g^s_n$ and $g_p^l$ are the free proton and neutron g factors, respectively, and $\left<\;\right>$ indicates a sum over all nucleons. The $\mu_{\rm IS}$, $\mu_{\rm IV}$ and $\left<s_3\right>$ each extracts a specific part of the magnetic moment operator and amplify small differences in experimental and theoretical values, and therefore are more sensitive to small changes in the magnetic moments of the mirror pair. The present $\mu(^{40}$Sc) was combined with the existing magnetic moment of $\mu(^{40}$K) \cite{sah74}, and $\mu_{\rm IS}$, $\mu_{\rm IV}$ and $\left<s_3\right>$ were obtained, which are summarized in Table \ref{tab:summary} and Figure \ref{fig:result} together with the shell model and IMSRG calculations. 

Good agreements were achieved for $\mu_{\rm IS}$ and $\mu_{\rm IV}$ between experiment and shell model calculations with the empirical shell model values evaluated using the additivity rule. The $\left<\sigma_3\right> $ has large uncertainty due to the uncertainty in the present $\mu$($^{40}$Sc), and no further discussion is given here. The IMSRG calculations show similarly good agreement except the NNLOgo(394) interaction. It can be seen that $\mu_{\rm IS}$ is insensitive to the different set of calculations, including the single particle value, and show only slight variation within the experimental uncertainties. On the other hand, calculated values of $\mu_{\rm IV}$ deviate more, and dominate the discrepancy of the magnetic moment from experimental value. 

\section{Summary}
\label{sec:summary}
Bunched-beam collinear laser spectroscopy was performed to determine electromagnetic moments of the 4$^-$ ground state in the odd-odd $^{40}$Sc nucleus, occurring at the proton drip line. The hyperfine structure of the $^3\text{D}_2\,\leftrightarrow\,^3\text{F}^\circ_3$ transition in the singly-charged $^{40}$Sc was measured, and the magnetic-dipole and electric quadrupole hyperfine coupling constants were deduced. Magnetic moments obtained by the additivity rule using empirical magnetic moments of neighboring odd-even nuclei and shell model corrections well reproduce the experimental magnetic moment of $\mu(^{40}$Sc), and the mirror nucleus $\mu(^{40}$K). The ab-initio multi-shell IMSRG calculations were also performed for the magnetic moments, which show good agreement with experimental value except the NNLOgo(394) interaction. However, the good agreement should be further confirmed with an inclusion of mesonic contribution, which is not included in the present calculations. The isoscalar and isovector magnetic moments were deduced with the known magnetic moment of mirror $^{40}$K nucleus. The $\mu_{\rm IS}$ is insensitive to the calculations employed in the present study, but the $\mu_{\rm IV}$ deviates more and dominates the departure from experimental magnetic moments. It is noted that the shell model calculations for the $A$ = 39 and 41 doubly-magic plus or minus one nucleon nuclei deviate from their experimental values, which indicates missing nucleon excitation across the $sd$ and $fp$ major shells. The good agreement of the empirical shell model calculations with the present $^{40}$Sc magnetic moment and the analysis of the additivity rule can be regarded that the additivity rule is still applicable for such systems, where cross shell excitation has considerable contribution to magnetic moments. The electric quadrupole moment of $^{40}$Sc was deduced from the present hyperfine coupling constant, but no further discussion was given due to large statistical uncertainty. 


%

\end{document}